\documentclass{article}

\usepackage{arxiv}

\usepackage[utf8]{inputenc} 
\usepackage[T1]{fontenc}    
\usepackage{hyperref}       
\usepackage{url}            
\usepackage{booktabs}       
\usepackage{amsfonts}       
\usepackage{nicefrac}       
\usepackage{microtype}      
\usepackage{lipsum}		
\usepackage{graphicx}
\usepackage{natbib}
\usepackage{doi}

\title{Understanding Impact of Angle in Urban Transportation}

\author{{Kota Nagasaki} \\
	Department of Civil and Environmental Engineering\\
	Tokyo Institute of Technology\\
	2-12-1-M1-13, Ookayama, Meguro, Tokyo \\
	\texttt{nagasaki.k.ab@m.titech.ac.jp} \\
	\And
	{Toru Seo} \\
        Department of Civil and Environmental Engineering\\
	Tokyo Institute of Technology\\
	2-12-1-M1-13, Ookayama, Meguro, Tokyo \\
	\texttt{seo.t.aa@m.titech.ac.jp} \\
}

\renewcommand{\shorttitle}{\textit{arXiv} Template}

\hypersetup{
pdftitle={A template for the arxiv style},
pdfsubject={q-bio.NC, q-bio.QM},
pdfauthor={David S.~Hippocampus, Elias D.~Striatum},
pdfkeywords={First keyword, Second keyword, More},
}

\begin{document}
\maketitle

\begin{abstract}
Traffic congestion at urban-scale levels occurs when road network supply is insufficient compared with demand. Therefore, the relationship between supply and demand has been extensively investigated in the literature.	Especially the impact of a network's topology (i.e., connectivity) on traffic congestion has been widely studied.	On the other hand, only a few studies analyze the relationship between the physical shape of the road network (i.e., morphology), demand, and congestion. For example, angular indicators associated with links and trips can be utilized to characterize such morphology. However, statistical analysis of angular indicators requires a specialized methodology called Directional Statistics, and how to apply Directional Statistics to traffic data is not obvious. Here, this study develops a descriptive model of traffic congestion based on Directional Statistics. It describes direction-dependent congestion levels using angular distributions that characterize the shape of road network and travel demand. Specifically, the shape of road network is characterized by the distribution of angles of each road link in a specific region.	Likewise, the travel demand is characterized by the distribution of angles of each trip in the region.	Then, a statistical model that describes travel time for a trip with arbitrary direction by these two angular distribution is constructed by paying special attention to the fact that roads and demands from all angles would affect congestion. A simple estimation method for the proposed model is also presented. Finally, the model was estimated using actual data in Tokyo, and the relationship between the characteristics of the shape of the road network, the direction of demand, and congestion is analyzed from the new perspective of angles.
\end{abstract}

\keywords{Angular indicator \and Directional statistics \and Network analysis \and Congestion model}

\section{Introduction}
Traffic congestion occurs when the supply of road network is insufficient compared to the demand at urban-scale levels.
Therefore, the relationship between supply and demand has been extensively investigated in the literature.
Specifically, the impact of the network's topology (i.e., connectivity) on traffic congestion has been widely studied \citep{sheffi1985urban, yang1998models, yin2001assessing, farahani2013review, rempe2016spatio, sun2018role}. 
Alternatively, simpler area-based approaches that focus on aggregated traffic behavior over a wide area have also been employed, such as the continuous modeling approach \citep{beckmann1952continuous, sasaki1990user, wong1998multi} and the network or macroscopic fundamental diagram \citep{mahmassani1984investigation, williams1987urban, daganzo2008analytical, ji2014empirical, saeedmanesh2017dynamic}.

These topology- or area-based approaches have been highly successful in providing valuable insights into urban-level traffic congestion.
For example, it has been observed that roads in urban centers experience higher congestion as a result of their higher betweenness centrality.
Furthermore, the implementation of detour routes has been shown to reduce overall travel time. 
Additionally, the capability of throughput of demand depends on traffic volume (i.e., fundamental diagram) at the network level, which does not depend on demand patterns \citep{daganzo2008analytical}.

However, these approaches encounter challenges in explaining certain phenomena. 
For example, although it is possible to quantify the congestion levels of individual origin-destination (OD) demands, providing a comprehensive explanation of their sensitivity to the overall demand pattern proves difficult.
Moreover, it is challenging to clarify the relationships between congestion and the physical characteristics of the road network, such as whether it follows a grid pattern or not.

The shape of a road network characterized by the distribution of the {\it angles} or {\it orientation} of each road in the network, which can be called {\it morphology} of road networks, has not received much attention in congestion analysis.
The variable of ``angle'' appears in various aspects of urban transportation and congestion.
On the supply side, each road segment has its own angle, while on the demand side, each origin-destination (OD) pair associated with a traveler also possesses an angle.
Analyzing traffic congestion by specifically focusing on these angles could potentially offer insights and contribute to a better understanding of congestion dynamics.

An angle-based approach may offer certain advantages over topology- or area-based approaches.
For example, by representing travel demand using angles, it becomes easier to describe the relationship of that particular demand within the overall demand pattern.
This could potentially enable a more comprehensive explanation of congestion sensitivity to demand patterns.
Furthermore, it could facilitate describing the relationship between the physical shape of the road network, characterized by angular indicators, and congestion dynamics.
By considering angles, a deeper understanding of how the road network's geometry influences congestion could be attained.

One of the challenges in using angular indicators is that they have a fundamentally different nature compared to ordinary numerical variables such as distance or area.
Specifically, angles span a full circle of 360 degrees or $2\pi$ radians.
This means that the angular difference between 1 degree and 359 degrees is actually {\it smaller} than the difference between 1 degree and 180 degrees.
Consequently, traditional statistical methods, which are effective for analyzing numerical variables, cannot be directly applied to angular variables.
This poses a barrier to employing conventional statistical techniques, which have been successful in topology- or area-based approaches, for analyzing traffic congestion using angles.
To overcome this challenge, the field of statistics has developed a specialized branch known as {\it Directional Statistics} \citep{mardia2000directional}.
Directional Statistics offers methods and techniques specifically designed to handle angular data.
However, its application to transportation research, particularly in the context of traffic congestion analysis, remains limited.

There have been several studies that have utilized angular indicators to describe various aspects of road traffic. 
For example, \cite{boeing2019urban} and \cite{nagasakihksts} employed a rose diagram, which is a histogram for angular data, to depict the distribution of road orientations within a road network.
These studies focused on describing the physical shape of the road network (i.e., morphology) rather than the traditional approach of analyzing the network based on connectivity relationships between nodes (i.e., topology).
On the other hand, \cite{zhou2015understanding} analyzed massive bike-sharing data in Chicago by employing a rose diagram to visualize the direction of the trip.
In this study, each trip is described as the angle of a vector from the origin to the user's destination.
However, no studies have investigated traffic congestion considering angles in both the demand and supply sides. 

In this context, the aim of this study is to develop a novel descriptive model that captures the phenomenon of traffic congestion by considering two angular distributions: the direction of demand and the shape of the road network.
The proposed model utilizes Directional Statistical methodology to predict direction-dependent traffic congestion based on the angular distributions of supply and demand.
To validate the model, actual traffic data from Tokyo is utilized.
The contribution of this study can be summarized as follows:
\begin{itemize}
\item Introducing a methodology for predicting the performance of transportation systems based on angular indicators of traffic demand and road network characteristics.
\item Presenting a novel approach that integrates directional statistics, a field less explored in transportation research, into the study of transportation phenomena.
\item Through a case study, quantitatively revealing the relationships between the shape of the road network, traffic demand, and congestion, utilizing angular analysis.
\end{itemize}

The paper is organized as follows.
Section \ref{model} describes a congestion model with angular distribution and details of two distributions, the estimation process, and the expected property of the model.
Section \ref{case} describes a case study on the actual road network with the congestion data and the interpretation of the results.
Section \ref{conclu} concludes this paper.

\section{Model}
\label{model}

\subsection{Overview}

In our model, we aim to capture the congestion levels for travel in a specific direction $\theta\in[0,2\pi)$ by taking into account the angular distribution of demand and the road network.
Specifically, we focus on congestion during peak hours and assume that both the angular distribution of the road network and the demand pattern influence congestion.

Let $c(\theta)$ represent the degree of congestion for travel in the direction $\theta$ within a given area, $d(\theta)$ denote the angular distribution of demand, and $n(\theta)$ do the angular distribution of the road network.

Here, $c(\theta)$ is expected to be influenced by the demand and road network at angles different from $\theta$ itself. 
For example, high demand in the north-south direction or insufficient road network development in the north-south direction may result in congestion in the east-west direction.
Our aim is to build a model that accounts for such complexities.
However, as a pioneering study in congestion analysis using angles, this paper proposes a simplified model to demonstrate the concept.


\subsection{Congestion model}

To account for the influence of demand and road network characteristics on congestion levels $c(\theta)$ in different directions, we introduce two additional variables, $\phi\in[-\pi,\pi)$ and $\eta\in[-\pi,\pi)$.
These variables allow us to describe $c(\theta)$ in terms of the demand distribution at angles different from $\theta$, denoted by $d(\theta+\phi)$, and the road network characteristics at angles different from $\theta$, denoted by $n(\theta+\eta)$.

Furthermore, the impact of these factors on $c(\theta)$ is expected to depend on the angular differences, $\phi$, and $\eta$. When $\phi$ and $\eta$ are close to zero, the effect on $c(\theta)$ is anticipated to be significant. 
Conversely, when $\phi$ and $\eta$ are around $\pm\pi$, indicating opposite directions, the effect is expected to be smaller.
At $\pm\pi/2$, indicating orthogonal directions, there may be a special effect due to the presence of intersections at right angles.

To incorporate this influence into the model, we introduce functions $\alpha(\phi)$ and $\beta(\eta)$ that manipulate the degree of impact on $c(\theta)$ based on the values of $\phi$ and $\eta$.

By considering these variables and functions, our model aims to capture the complex relationship between congestion, demand distribution, road network characteristics, and the angular differences between them.

Based on the above, $c(\theta)$ can be described as
\begin{equation}
  c(\theta)=\int_{\phi}\int_{\eta}f\big(d(\theta+\phi)\alpha(\phi),n(\theta+\eta)\beta(\eta)\big).
\label{eq:intint}
\end{equation}
As the simplest case under this condition, assuming that $d(\theta)$ and $n(\theta)$ independently affect $c(\theta)$.
Then, Eq.\ (\ref{eq:intint}) can be converted to a linear combination of $d(\theta)$ terms and $n(\theta)$ terms as
\begin{equation}
  c(\theta)=\int_{\phi}d(\theta+\phi)\alpha(\phi)+\int_{\eta}n(\theta+\eta)\beta(\eta)+\gamma,
\label{eq:eq1}
\end{equation}
where $\gamma$ is a constant term.

Here, it is difficult to handle the $\alpha(\phi)$ and $\beta(\eta)$ because these variables are angles.
Therefore, the following transformation is performed to achieve linear regression with angle variables \citep{johnson1978some}.
\begin{equation}
\alpha(\phi)=\sum^{K}_{k=1}(\alpha_{\rm{c}\it{k}}\cos{k\phi}+\alpha_{\rm{s}\it{k}}\sin{k\phi}),\\
\beta(\eta)=\sum^{K}_{k=1}(\beta_{\rm{c}\it{k}}\cos{k\eta}+\beta_{\rm{s}\it{k}}\sin{k\eta}),
\label{eq:eq2}
\end{equation}
where $K$ is an external parameter, an integer that determines up to which degrees the distributions are combined.
As with the Fourier transform, employing the terms with higher degrees $K$ allows us to obtain a finer distribution of $\alpha(\phi)$ and $\beta(\eta)$.

Substituting Eq.\ (\ref{eq:eq2}) into Eq.\ (\ref{eq:eq1}) gives
\begin{equation}
c(\theta)=\int_{\phi}\Big\{d(\theta+\phi)\sum^{K}_{k=1}(\alpha_{\rm{c}\it{k}}\cos{k\phi}+\alpha_{\rm{s}\it{k}}\sin{k\phi})\Big\}+\\
\int_{\eta}\Big\{n(\theta+\eta)\sum^{K}_{k=1}(\beta_{\rm{c}\it{k}}\cos{k\eta}+\beta_{\rm{s}\it{k}}\sin{k\eta})\Big\}+\gamma.
\label{eq:modelcon}    
\end{equation}

If each distribution is discrete rather than continuous, then
\begin{equation}
  c(\theta)=\sum_{\phi\in\Phi}\Big\{d(\theta+\phi)\sum^{K}_{k=1}(\alpha_{\rm{c}\it{k}}\cos{k\phi}+\alpha_{\rm{s}\it{k}}\sin{k\phi})\Big\}+\\
  \sum_{\eta\in H}\Big\{n(\theta+\eta)\sum^{K}_{k=1}(\beta_{\rm{c}\it{k}}\cos{k\eta}+\beta_{\rm{s}\it{k}}\sin{k\eta})\Big\}+\gamma.
\label{eq:modeldis}    
\end{equation}
where $\Phi$ and $H$ are the set of values of the difference between the representative point of each bin of the discretized angular distribution of demand $d$ and the road network $n$ and $\theta$, respectively.
The parameters to be estimated are $\alpha_{\rm{c}\it{k}}$, $\alpha_{\rm{s}\it{k}}$, $\beta_{\rm{c}\it{k}}$, $\beta_{\rm{s}\it{k}}$ and $\gamma$.
Note that this model is a linear regression model.
The estimation method is discussed later.

\begin{figure}[t]
	\centering
	\includegraphics[width=0.95\textwidth]{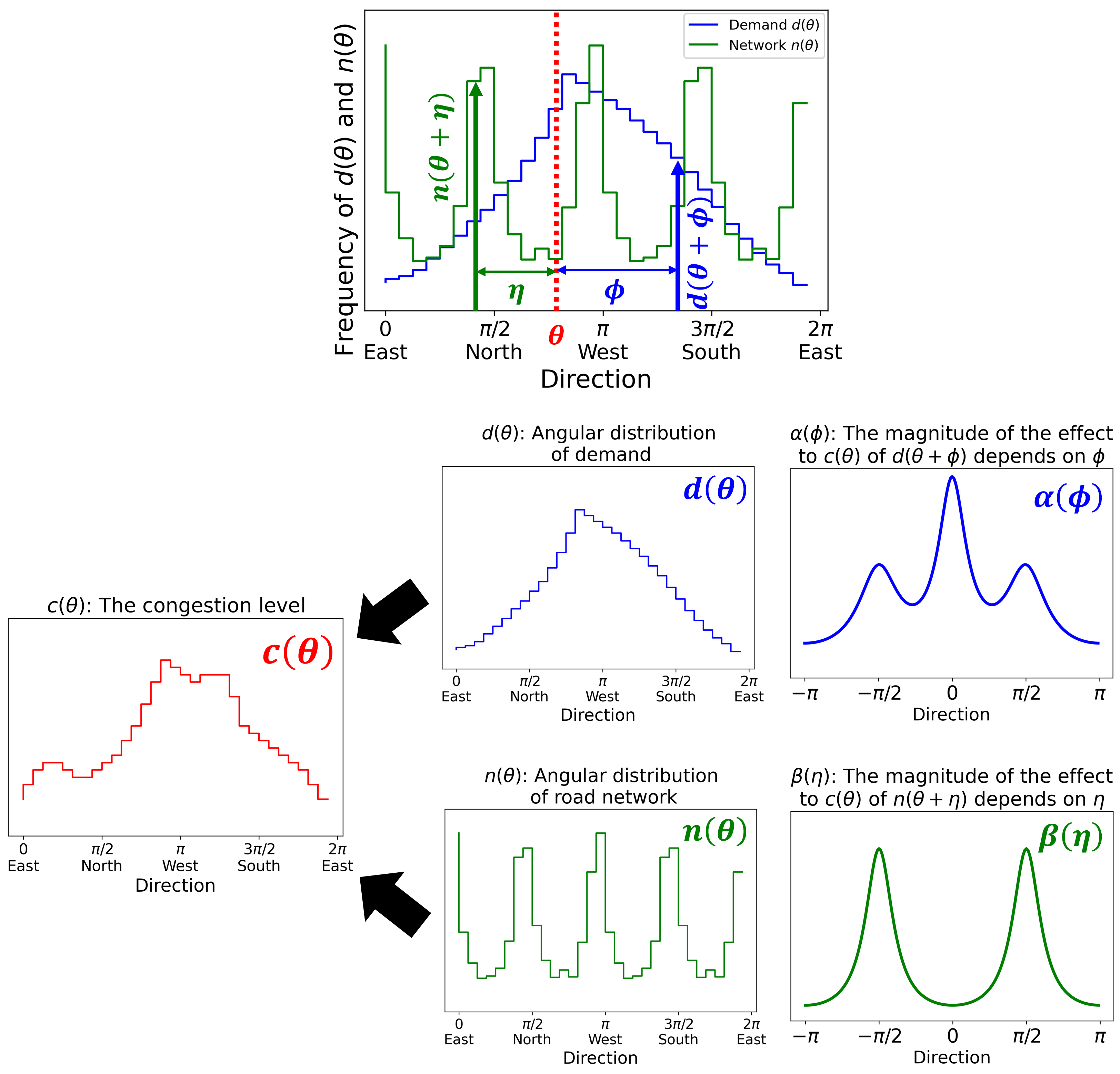}
	\caption{Conceptual diagram of the model.}
	\label{fig:flow}
\end{figure}

Fig.\ \ref{fig:flow} describes the conceptual diagram of the model.
In the proposed congestion model, $c(\theta)$ represents the congestion level, while $d(\theta)$ and $n(\theta)$ correspond to the angular distributions of demand and road network, respectively.

Considering that $c(\theta)$ can be influenced by the entire distribution of $d(\theta)$ and $n(\theta)$, the model incorporates the effects of these distributions based on the values of $\phi$ and $\eta$. To quantify this influence, functions $\alpha(\phi)$ and $\beta(\eta)$ are introduced, representing the degrees of influence depending on the values of $\phi$ and $\eta$.

Overall, this congestion model predicts $c(\theta)$ for each trip using the distributions of $d(\theta)$ and $n(\theta)$ within the target area as explanatory variables, along with $\alpha(\phi)$ and $\beta(\eta)$ as corresponding parameters. By considering these factors, the model aims to capture the complex relationship between congestion, demand distribution, road network characteristics, and the influence of angular differences.

\subsection{Specification}

The proposed model does not restrict the method of deriving $c(\theta)$, $d(\theta)$ and $n(\theta)$.
This section introduces an example of a method for deriving them.

\subsubsection{Congestion}

One commonly used indicator to assess the level of congestion in a specific direction is the average speed or its inverse, known as pace, of trips.
These indicators can be easily calculated using trajectory data, such as probe data.

\subsubsection{Demand}

The calculation of $d(\theta)$ relies on data containing origin and destination information. $d(\theta)$ is obtained by aggregating the directions from the origin to the destination for demand within the target area.
The origin and destination points can also be the coordinates of entering or exiting the target area rather than the origin and destination points of the entire trip.
Therefore, if trajectory data is used to derive $c(\theta)$, the data can be used as demand data.
However, note that the demand, in this case, is only the observed traffic volume, which is different from the actual travel demand.

\subsubsection{Network}

\begin{figure}[t]
	\centering
	\includegraphics[width=0.8\textwidth]{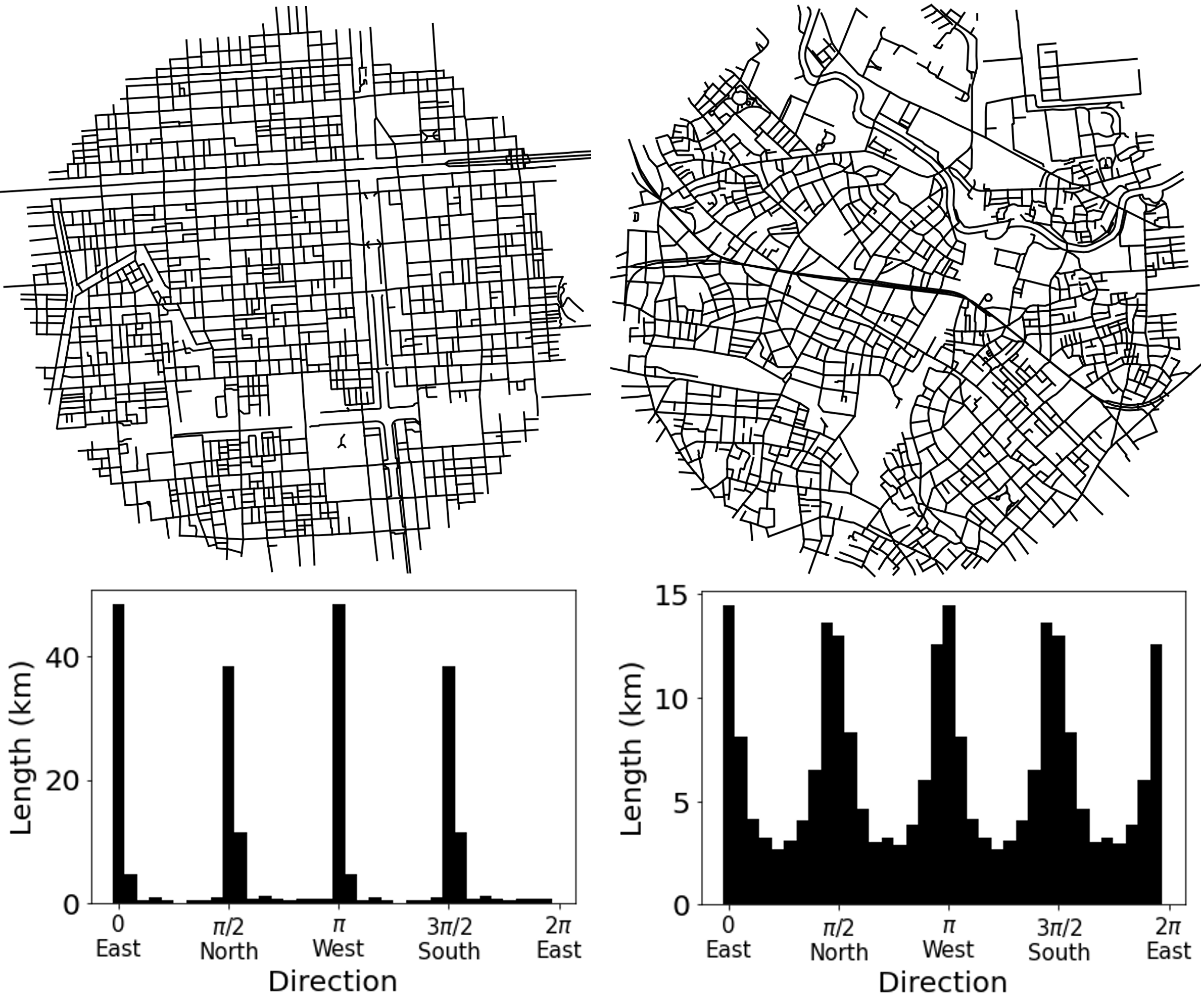}
	\caption{Example of $n(\theta)$ and corresponding road networks for a 1 km radius area.}
	\label{fig:ex_n}
\end{figure}

The angular distribution of the road network $n(\theta)$ is the aggregate of the angles that the roads face, called orientation, within the target area \citep{boeing2019urban, nagasakihksts}.
In this distribution, the peak is sharper when many roads are oriented in a particular direction, like a grid network of roads as shown in the left of Fig.\ \ref{fig:ex_n}.
When the road network is not a grid, the peaks are gradual, and the skirts are higher, as shown in the right of Fig.\ \ref{fig:ex_n}.
The large number of roads oriented in a particular direction can be regarded as a high capacity to serve the demand in that direction.
Therefore, this method is a suitable distribution for $n(\theta)$, the supply indicator in this analysis.

Note that these distributions can be transformed into continuous circular distributions easily by employing the distributions used in directional statistics and their estimation methods \citep{nagasaki2022traffic}.

\subsection{Estimation}

The parameters to be estimated in the model are $\alpha_{\rm{c}\it{k}}$, $\alpha_{\rm{s}\it{k}}$, $\beta_{\rm{c}\it{k}}$, $\beta_{\rm{s}\it{k}}$ and $\gamma$.
In addition, $K$ is an external parameter that determines the fine granularity of the estimated $\alpha(\phi)$ and $\beta(\eta)$ and needs to be determined a priori before the estimation.
A larger $K$ may provide a better estimate; however, it is not recommended because of the interpretability of the estimation results.

Eq.\ (\ref{eq:modelcon}) and Eq.\ (\ref{eq:modeldis}) are linear regression models and their parameters are obtained by the least squares method.
Eq.\ (\ref{eq:modeldis}) can be transformed as
\begin{equation}
c(\theta)=\sum^K_{k+1}\Big\{\alpha_{\rm{c}\it{k}}\sum_{\phi\in\Phi}d(\theta+\phi)\cos{k\phi}+\alpha_{\rm{s}\it{k}}\sum_{\phi\in\Phi}d(\theta+\phi)\sin{k\phi}\\
+\beta_{\rm{c}\it{k}}\sum_{\eta\in H}n(\theta+\eta)\cos{k\eta}+\beta_{\rm{s}\it{k}}\sum_{\eta\in H}n(\theta+\eta)\sin{k\eta}\Big\}+\gamma.
\label{eq:paracoe}    
\end{equation}
The parameters can be estimated by performing a linear regression with the given $d(\theta)$ and $n(\theta)$ and the trigonometric values substituted, and the congestion indicator $c(\theta)$ as the objective function.

In addition, by substituting the estimated $\alpha_{\rm{c}\it{k}}$, $\alpha_{\rm{s}\it{k}}$, $\beta_{\rm{c}\it{k}}$ and $\beta_{\rm{s}\it{k}}$ into Eq.\ (\ref{eq:eq2}), the distributions $\alpha(\phi)$ and $\beta(\eta)$ of the magnitude of the effect of $d(\theta+\phi)$ and $n(\theta+\eta)$ on $c(\theta)$ by the values of $\phi$ and $\eta$ can be obtained.
The shape of the distributions provides a quantitative indication of the magnitude to which demand in a given direction is affected by other direction demand and road, e.g., orthogonal.

Note that if the distribution of the road network is point symmetric like \cite{boeing2019urban}, i.e., all roads are aggregated in both directions, then the distribution of $\beta(\eta)$ is also point symmetric.
Therefore, the terms $\beta_{\rm{c}\it{k}}$ and $\beta_{\rm{s}\it{k}}$ can be deleted in the case $k$ is odd.

\subsection{Discussion}

This model is a simple congestion model using the angular distribution of the demand and the road network.
The parameters to be estimated are $\alpha_{\rm{c}\it{k}}$, $\alpha_{\rm{s}\it{k}}$, $\beta_{\rm{c}\it{k}}$, $\beta_{\rm{s}\it{k}}$ for each degree $k$.
However, the interpretation of the model is enhanced by examining the shape of the $\alpha(\phi)$ and $\beta(\eta)$ distributions formed by these parameters.

Because various factors cause the congestion phenomenon, the distribution of $\alpha(\phi)$ and $\beta(\eta)$ could be complicated and multimodal.
Some expected results are: (1) $\alpha(0)$ is expected to be positive because much demand in the same direction is expected to cause more congestion in that direction.
(2) $\beta(0)$ is expected to be negative because the lack of road construction in a given direction is expected to cause congestion in demand toward that direction.

\section{Case Study}
\label{case}

\subsection{Data and variables}

The case study was conducted in the vicinity of the Tokyo metropolitan area. The data used to derive $c(\theta)$ and $d(\theta)$ were obtained from vehicle trajectory data collected between July 12, 2021 (Monday) and July 16, 2021 (Friday).

For each trip, the start and end times are determined when the probe car enters or exits the target area, or when the car commences or concludes a trip within the area. The duration of the trip is calculated as the difference between these times, resulting in the trip time (in seconds). $c(\theta)$ is computed as the ratio of travel time to the distance covered (in kilometers) for each trip. On the other hand, $d(\theta)$ represents the aggregate of the angles formed between the origin and destination of each trip.

To ensure data quality, the top 10$\%$ of $c(\theta)$ values were excluded, as some trips were deemed ongoing despite the vehicle staying in one location for an extended period. Similarly, the bottom 5$\%$ of $c(\theta)$ values were omitted due to their apparent high speeds.

The network distribution $n(\theta)$ was derived using OpenStreetMap. The analysis focused on major roads classified as "motorway", "trunk", "primary", and "secondary."

Both $d(\theta)$ and $n(\theta)$ were normalized frequencies represented as discrete histograms with 32 bins. The parameter $K$ was set to 8 for improved interpretability.

\begin{figure}[t]
	\centering
	\includegraphics[width=0.9\textwidth]{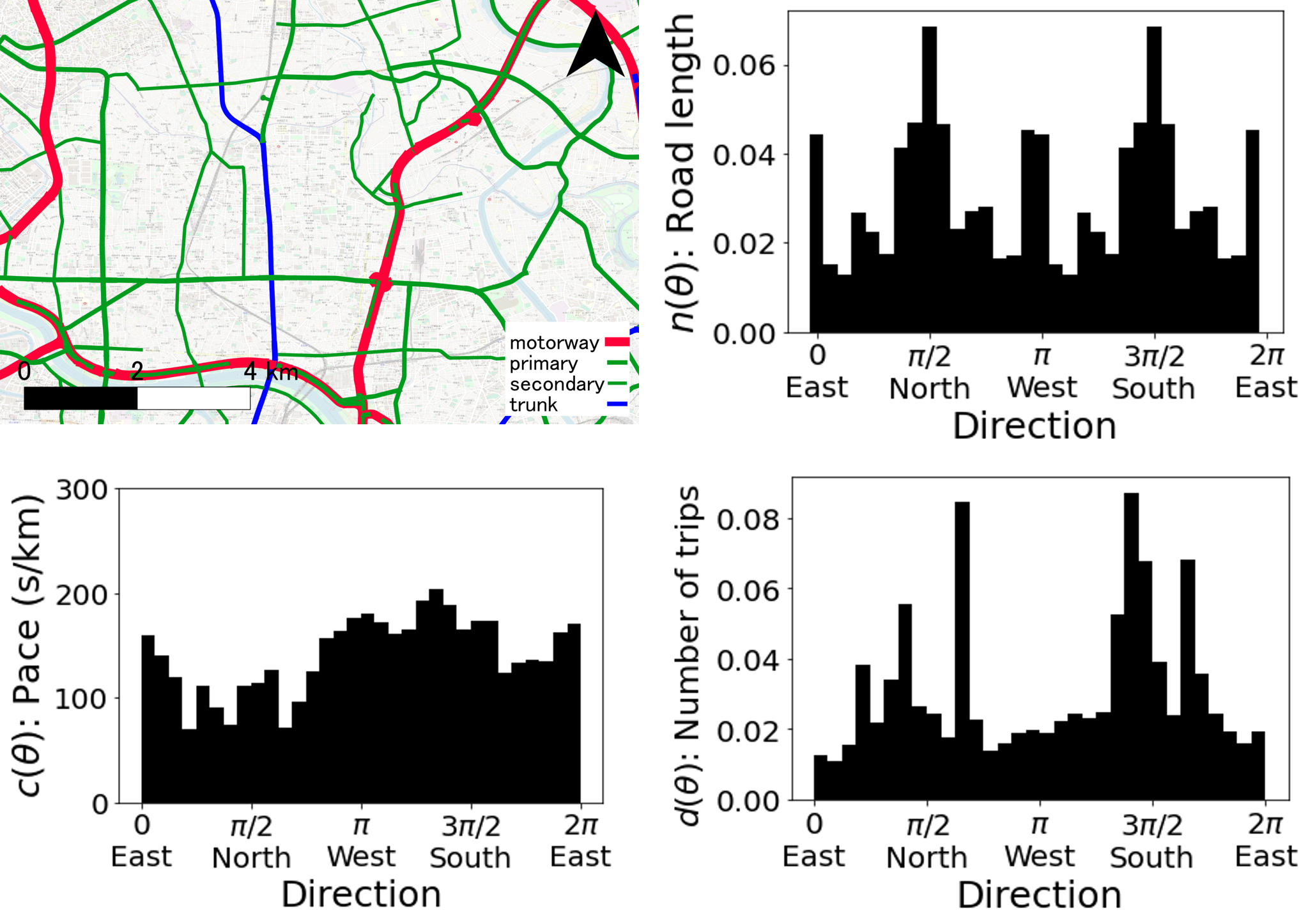}
	\caption{The maps of the road networks and the distributions (upper left) of $c(\theta)$ (bottom left), $d(\theta)$ (bottom right), and $n(\theta)$ (upper right) for Case 1 (North of the center of Tokyo in morning). Maps exported from OpenStreetMap.}
	\label{fig:case1}
\end{figure}

\begin{figure}[t]
	\centering
	\includegraphics[width=0.9\textwidth]{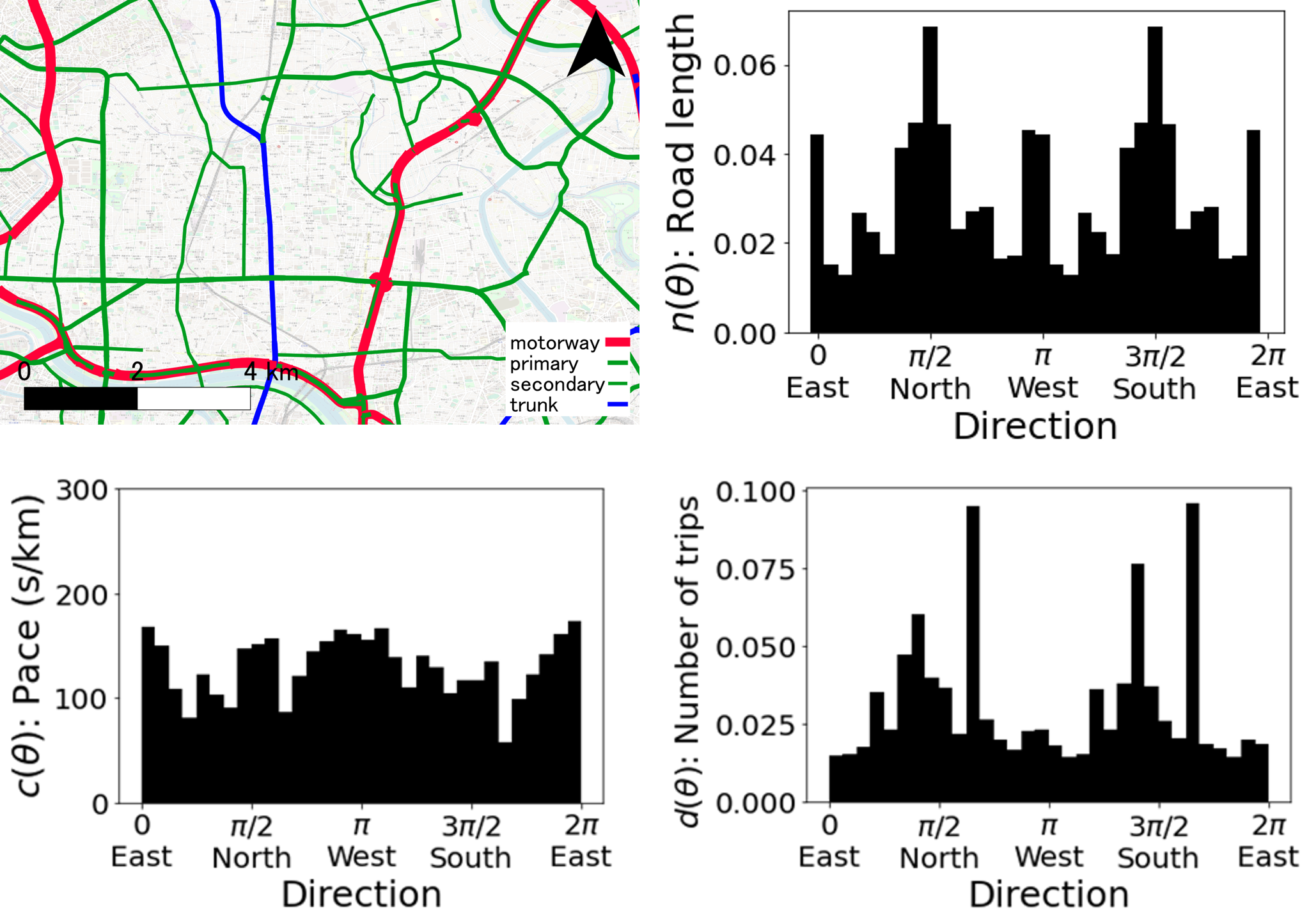}
	\caption{The maps of the road networks and the distributions (upper left) of $c(\theta)$ (bottom left), $d(\theta)$ (bottom right), and $n(\theta)$ (upper right) for Case 2 (North of the center of Tokyo in evening). Maps exported from OpenStreetMap.}
	\label{fig:case2}
\end{figure}

\begin{figure}[t]
	\centering
	\includegraphics[width=0.9\textwidth]{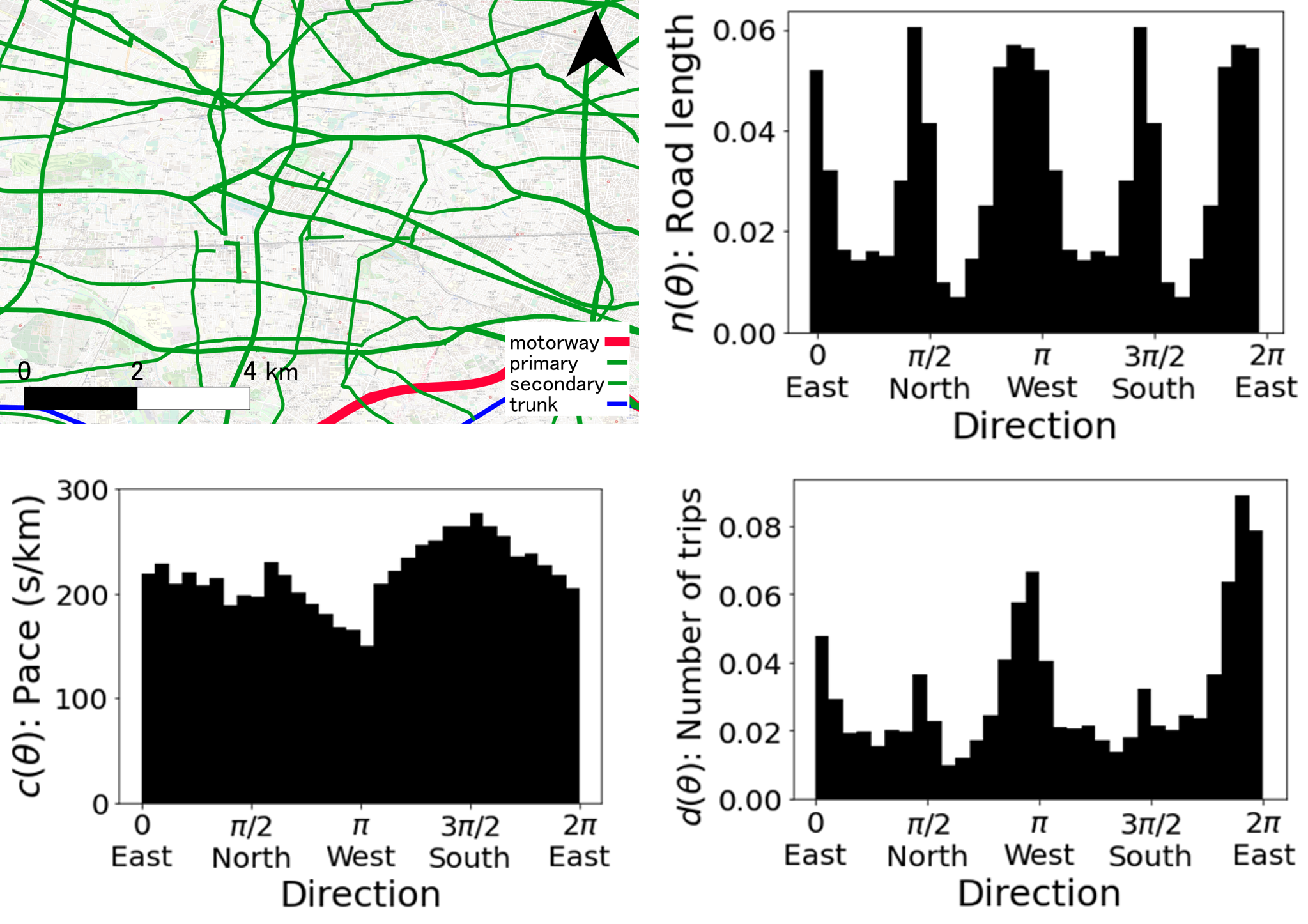}
	\caption{The maps of the road networks and the distributions (upper left) of $c(\theta)$ (bottom left), $d(\theta)$ (bottom right), and $n(\theta)$ (upper right) for Case 3 (West of the center of Tokyo in morning). Maps exported from OpenStreetMap.}
	\label{fig:case3}
\end{figure}

Three case studies were conducted.
Case 1 and 2 are the 8:00 a.m. and 6:00 p.m. trip data for a 10 km square area located north of the center of Tokyo.
Case 3 is the 8:00 a.m. trip data for a 10 km square area located west of the center of Tokyo.
The map of the road network and the distributions of $c(\theta)$, $d(\theta)$, and $n(\theta)$ for each case are shown in Fig.\ \ref{fig:case1}, Fig.\ \ref{fig:case2} and Fig.\ \ref{fig:case3}.

For Case 1 and Case 2, demand is higher in the south direction toward the center of Tokyo and in the opposite direction to the north.
In addition, $c(\theta)$ in these directions is small because of the highways.
$n(\theta)$ has a significant peak in the north-south direction due to the large number of roads in that direction.
For Case 3, demand is higher in the east direction toward the center of Tokyo and in the opposite direction to the west.
However, $c(\theta)$ is generally large because there are no highways in the area.
$n(\theta)$ has a significant peak in the east-west direction due to the large number of roads in that direction.

\subsection{Estimation results}

\begin{figure}[t]
	\centering
	\includegraphics[width=0.9\textwidth]{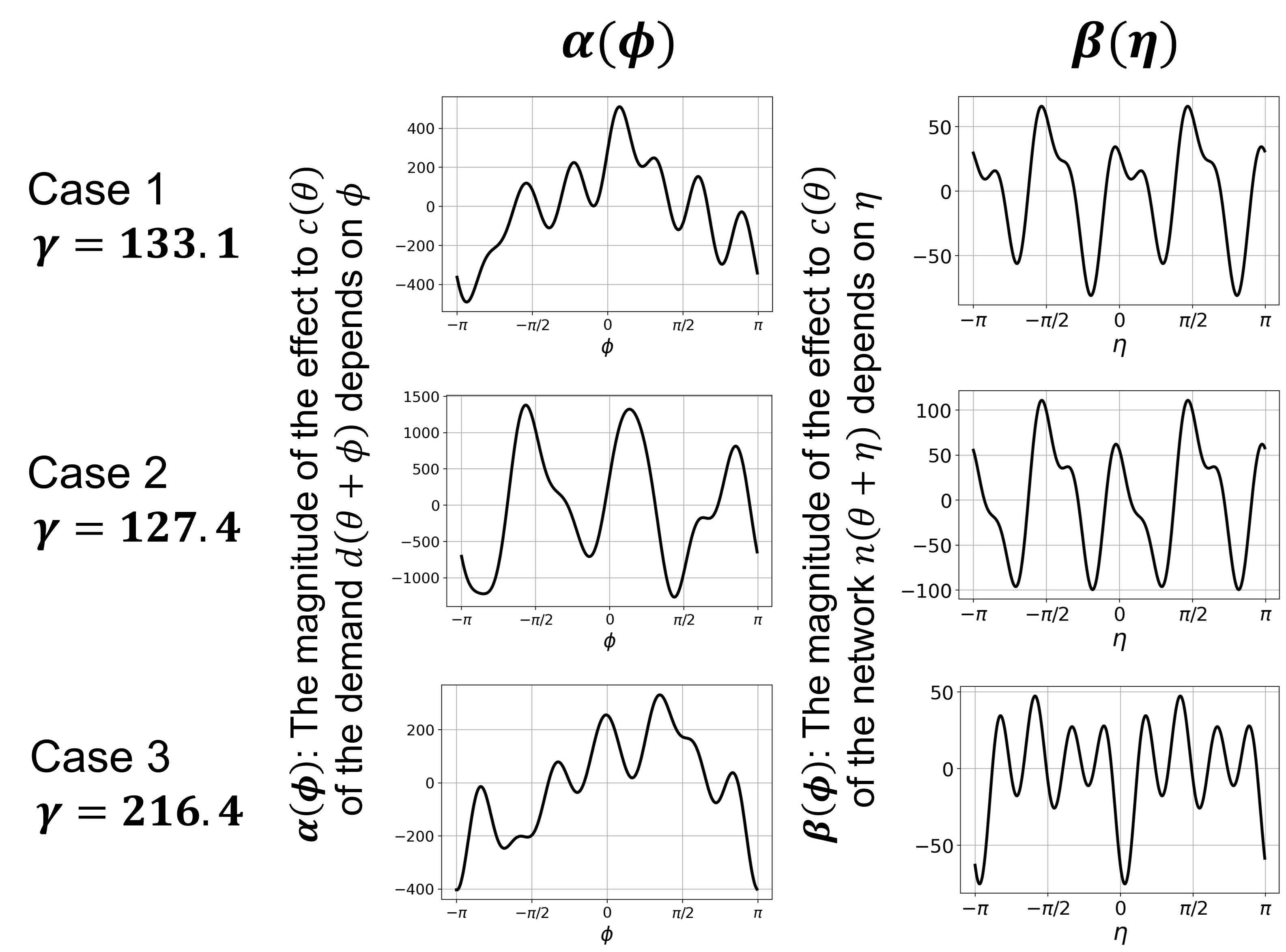}
	\caption{The distribution of $\alpha(\phi)$ and $\beta(\eta)$ estimated in each case.}
	\label{fig:eachdist}
\end{figure}

The distributions of $\alpha(\phi)$ and $\beta(\eta)$ estimated in each case are shown in Fig.\ \ref{fig:eachdist}.
The value of $R^2$, $F$-statistic, Prob($F$-statistic) are shown in Table \ref{tab:case1}, Table \ref{tab:case2} and Table \ref{tab:case3}.
The value of parameters for each case are shown in Appendix.
Note that $\alpha(\phi)$ and $\beta(\eta)$ are calculated by only the 5$\%$ significant parameters.
In each case, the number of parameters that are 5$\%$ significant out of a total of 16 $\alpha_{\rm{c}\it{k}}$ and $\alpha_{\rm{s}\it{k}}$ is 14, 13 and 10, respectively.
In addition, the number of parameters that are 5$\%$ significant out of a total of 8 $\beta_{\rm{c}\it{k}}$ and $\beta_{\rm{s}\it{k}}$ is 5, 8 and 6, respectively.

\begin{table}[ht]
	\centering
	\caption{Case 1 results.}
	\label{tab:case1}
 	\begin{tabular}{cr}
		\hline
        & Value\\
		\hline
        Number of samples & 10969\\
        $R^2$ & 0.307\\
        $F$-statistic & 302.8\\
        Prob($F$-statistic) & 0.000\\
        \hline
	\end{tabular}
\end{table}

\begin{table}[ht]
	\centering
	\caption{Case 2 results.}
	\label{tab:case2}
 	\begin{tabular}{cr}
		\hline
        & Value\\
		\hline
        Number of samples & 9557\\
        $R^2$ & 0.174\\
        $F$-statistic & 126.0\\
        Prob($F$-statistic) & 0.000\\
        \hline
	\end{tabular}
\end{table}

\begin{table}[ht]
	\centering
	\caption{Case 3 results.}
	\label{tab:case3}
 	\begin{tabular}{cr}
		\hline
        & Value\\
		\hline
        Number of samples & 3517\\
        $R^2$ & 0.206\\
        $F$-statistic & 56.6\\
        Prob($F$-statistic) & 0.000\\
        \hline
	\end{tabular}
\end{table}

According to the value of Prob($F$-statistic), all models are significant, and each parameter, including the constant term $\gamma$, is also working significantly.

\subsection{Discussion}

The distribution depicted in Fig.\ \ref{fig:eachdist} illustrates the impact of $d(\theta+\phi)$ and $n(\theta+\eta)$ values on $c(\theta)$.
In Case 1, for instance, if $d(\theta)$ is 0.1, $c(\theta)$ increases by approximately 25, whereas if $d(\theta+\pi)$ is 0.1, $c(\theta)$ decreases by approximately 40.
However, it is unlikely that demand in the opposite direction would decrease congestion.
On the contrary, it is reasonable to assume that such demand would have no impact on congestion, rather than reducing it.
This discrepancy arises from the possibility of $\alpha(\phi)$ having negative values, inherent to the characteristics of the model.
Therefore, it is natural to interpret $\alpha(\phi)$ not based on its value but rather by its incremental change from the smallest value.

Across all cases, $\alpha(\phi)$ is positive and exhibits significant magnitudes near $\phi=0$.
This outcome aligns with expectations, as it indicates that a high volume of demand in the same direction leads to increased congestion in that specific direction.
Furthermore, $\alpha(\phi)$ is significantly small around $\phi=\pi$, which is also expected, as demand in the opposite direction is unlikely to affect congestion.

However, the overall shape of $\alpha(\phi)$ differs between Case 1 and Case 2, despite the same target area.
Although the distributions of the two $d(\theta)$ in Fig.\ \ref{fig:case1} and Fig.\ \ref{fig:case2} are not significantly different, the shape of $c(\theta)$ in the southern direction differs. 
This discrepancy can be attributed to the heavy traffic towards central Tokyo during the morning peak period, resulting in congestion even for travelers using highways.
Conversely, congestion is reduced in the evening due to less crowded highways and faster travel speeds.
These factors contribute to the differing shapes of $\alpha(\phi)$ in Case 1 and Case 2.

In Cases 1 and 3, $\alpha(\phi)$ generally exhibits high values near $\phi=0$, contrasting with the highly oscillatory $\alpha(\phi)$ observed in Case 2.
This discrepancy can be attributed to the distribution of demand in directions other than the peak, as seen in $d(\theta)$ for Cases 1 and 3.
Consequently, congestion is more likely to occur when demand is distributed over a wide angular range, such as during the morning period.

Additionally, $\beta(0)$ is negative in Case 3, whereas it is positive in Case 1 and Case 2.
This contradicts initial expectations, as it suggests that congestion occurs when the road network aligns with travel demand.
This discrepancy may caused by the differences between the diagonal highway and the many east-west, north-south, and west-oriented roads.
While $c(\theta)$ in the diagonal direction is smaller due to the presence of the highway, $n(\theta)$ in that direction is also lower compared to other directions.
Hence, $\beta(0)$ is positive in Cases 1 and 2.

Furthermore, when comparing $\beta(\eta)$ in Case 1 and Case 2, the impact of the road network on congestion exhibits similar shapes during both morning and evening periods.
This finding aligns with the results of \cite{daganzo2008analytical}, indicating that the road network's influence on congestion follows a similar pattern regardless of demand patterns.

In Cases 1 and 2, $\beta(\eta)$ exhibits four peaks, which could be attributed to the road network distribution $n(\theta)$ also having four peaks.
However, Case 3 displays four peaks in $n(\theta)$, but $\beta(\eta)$ exhibits eight peaks.
The interpretation of this discrepancy remains elusive at present.

Furthermore, the magnitudes of $\alpha(\phi)$ and $\beta(\eta)$ differ by approximately tenfold.This indicates that demand has a stronger influence on congestion than the shape of the road network.

\section{Conclusion}
\label{conclu}

This study developed a novel descriptive model that describes the traffic congestion phenomenon using two angular distributions: the shape of the road network and the direction of demand.
The proposed model predicts direction-dependent traffic congestion using the two angular distributions using the Directional Statistical methodology.
In concrete, the model describes congestion using the angular distribution of demand, described by the direction of origin and destination of trips in the target area, and the angular distribution of the road network, described by the direction in which all roads are oriented.
To incorporate the two angular distributions into the model, the shape of the angular distribution is transformed into a cosine term and a sine term using directional statistics.
This made it possible to predict congestion with a simple linear regression model with angular indicators.
Furthermore, a case study of the proposed model was conducted in the Tokyo Metropolitan Area.
The model worked significantly in the case study.

The finding of this study can be summarized as follows:
\begin{itemize}
\item The entire angular distribution of both demand and the road network affects congestion in a particular direction.
\item Congestion during the morning hours is affected by a wider angle of demand.
\end{itemize}

To the authors' knowledge, this is the first attempt to describe congestion phenomena using angular indicators with Directional Statistics.
Therefore, there is still room for further development.
First, the connection to well-established concepts in transportation researches, such as Macroscopic Fundamental Diagram \citep{daganzo2008analytical} would be worth investigating.
Second, a dynamic model that introduces a time term could be developed by extending this result.
Time loops in 24 hours so that it can be described as angles \citep{nagasaki2022traffic}.
The data with two angular axes, demand and time, can be described by a torus, which is a doughnut-shaped surface.
The establishment of more advanced models using the latest findings in directional statistics is one of the development directions of this research.

\section*{Acknowledgement}
The data was freely provided by Tokyo Metropolitan Government ("offering of several datasets on mobility and transportation in Tokyo 2020 Games"),
This research was partially supported by JSPS Kakenhi 22J11294.

\bibliographystyle{unsrtnat}
\bibliography{references}  

\appendix
\section{Parameters of Each Case}

Estimated parameters for each model are shown in Table \ref{tab:case1para}, Table \ref{tab:case2para} and Table \ref{tab:case3para}.
``$^{*}$'' means 5$\%$ significant.

\begin{table}[ht]
	\centering
	\caption{Estimated parameters of Case 1.}
	\label{tab:case1para}
 	\begin{tabular}{crrr|crrr}
		\hline
		& Coefficient & Std. err. & $t$ value & & Coefficient & Std. err. & $t$ value\\
		\hline
$\gamma$ & 133.08 & 0.51 & $^{*}$258.96 & & & & \\
$\alpha_{\rm{c}1}$ & 263.11 & 5.63 & $^{*}$46.76 & $\alpha_{\rm{s}1}$ & 52.58 & 6.86 & $^{*}$7.66 \\
$\alpha_{\rm{c}2}$ & -18.40 & 2.01 & $^{*}$-9.14 & $\alpha_{\rm{s}2}$ & 21.84 & 1.89 & $^{*}$11.58 \\
$\alpha_{\rm{c}3}$ & 54.03 & 12.33 & $^{*}$4.38 & $\alpha_{\rm{s}3}$ & 85.40 & 11.56 & $^{*}$7.39 \\
$\alpha_{\rm{c}4}$ & 1.47 & 0.46 & $^{*}$3.23 & $\alpha_{\rm{s}4}$ & -6.62 & 0.43 & $^{*}$-15.24 \\
$\alpha_{\rm{c}5}$ & -54.29 & 6.61 & $^{*}$-8.21 & $\alpha_{\rm{s}5}$ & 55.51 & 6.81 & $^{*}$8.15 \\
$\alpha_{\rm{c}6}$ & -3.14 & 4.38 & -0.72 & $\alpha_{\rm{s}6}$ & 6.27 & 5.02 & 1.25 \\
$\alpha_{\rm{c}7}$ & 61.55 & 14.11 & $^{*}$4.36 & $\alpha_{\rm{s}7}$ & 108.00 & 13.98 & $^{*}$7.72 \\
$\alpha_{\rm{c}8}$ & -21.25 & 2.01 & $^{*}$-10.58 & $\alpha_{\rm{s}8}$ & 17.16 & 2.06 & $^{*}$8.32 \\
$\beta_{\rm{c}2}$ & -14.4 & 1.17 & $^{*}$-12.30 & $\beta_{\rm{s}2}$ & 8.22 & 1.09 & $^{*}$7.52 \\
$\beta_{\rm{c}4}$ & 43.93 & 3.14 & $^{*}$14.01 & $\beta_{\rm{s}4}$ & 18.34 & 3.12 & $^{*}$5.87 \\
$\beta_{\rm{c}6}$ & 1.54 & 1.13 & 1.36 & $\beta_{\rm{s}6}$ & -0.54 & 1.06 & -0.50 \\
$\beta_{\rm{c}8}$ & -0.49 & 1.75 & -0.28 & $\beta_{\rm{s}8}$ & -23.37 & 1.73 & $^{*}$-13.51 \\
		\hline
	\end{tabular}
\end{table}

\begin{table}[ht]
	\centering
	\caption{Estimated parameters of Case 2.}
	\label{tab:case2para}
 	\begin{tabular}{crrr|crrr}
		\hline
		& Coefficient & Std. err. & $t$ value & & Coefficient & Std. err. & $t$ value\\
		\hline
$\gamma$ & 127.39 & 0.64 & $^{*}$198.93 & & & & \\
$\alpha_{\rm{c}1}$ & 298.21 & 35.85 & $^{*}$8.32 & $\alpha_{\rm{s}1}$ & -158.73 & 33.74 & $^{*}$-4.70 \\
$\alpha_{\rm{c}2}$ & -49.84 & 2.53 & $^{*}$-19.70 & $\alpha_{\rm{s}2}$ & -3.55 & 2.38 & -1.49 \\
$\alpha_{\rm{c}3}$ & 263.20 & 88.45 & $^{*}$2.98 & $\alpha_{\rm{s}3}$ & 979.97 & 80.58 & $^{*}$12.16 \\
$\alpha_{\rm{c}4}$ & -6.82 & 0.36 & $^{*}$-19.09 & $\alpha_{\rm{s}4}$ & 0.33 & 0.39 & 0.85 \\
$\alpha_{\rm{c}5}$ & -196.84 & 54.38 & $^{*}$-3.62 & $\alpha_{\rm{s}5}$ & 162.29 & 54.16 & $^{*}$3.00 \\
$\alpha_{\rm{c}6}$ & -48.58 & 4.80 & $^{*}$-10.12 & $\alpha_{\rm{s}6}$ & 19.50 & 5.39 & $^{*}$3.62 \\
$\alpha_{\rm{c}7}$ & 193.99 & 60.28 & $^{*}$3.22 & $\alpha_{\rm{s}7}$ & 14.78 & 64.19 & 0.23 \\
$\alpha_{\rm{c}8}$ & -39.92 & 2.44 & $^{*}$-16.33 & $\alpha_{\rm{s}8}$ & 22.32 & 2.47 & $^{*}$9.02 \\
$\beta_{\rm{c}2}$ & -28.49 & 1.53 & $^{*}$-18.65 & $\beta_{\rm{s}2}$ & -9.76 & 1.43 & $^{*}$-6.82 \\
$\beta_{\rm{c}4}$ & 69.37 & 3.96 & $^{*}$17.52 & $\beta_{\rm{s}4}$ & 17.30 & 3.89 & $^{*}$4.45 \\
$\beta_{\rm{c}6}$ & 6.89 & 1.04 & $^{*}$6.63 & $\beta_{\rm{s}6}$ & 7.89 & 1.00 & $^{*}$7.86 \\
$\beta_{\rm{c}8}$ & 7.61 & 1.83 & $^{*}$4.15 & $\beta_{\rm{s}8}$ & -32.95 & 1.80 & $^{*}$-18.28 \\
		\hline
	\end{tabular}
\end{table}

\begin{table}[ht]
	\centering
	\caption{Estimated parameters of Case 3.}
	\label{tab:case3para}
 	\begin{tabular}{crrr|crrr}
		\hline
		& Coefficient & Std. err. & $t$ value & & Coefficient & Std. err. & $t$ value\\
		\hline
$\gamma$ & 216.41 & 0.81 & $^{*}$268.16 & & & & \\
$\alpha_{\rm{c}1}$ & 161.30 & 9.51 & $^{*}$16.97 & $\alpha_{\rm{s}1}$ & 138.52 & 11.52 & $^{*}$12.02 \\
$\alpha_{\rm{c}2}$ & -30.96 & 2.67 & $^{*}$-11.61 & $\alpha_{\rm{s}2}$ & 1.29 & 3.13 & 0.41 \\
$\alpha_{\rm{c}3}$ & -2.75 & 18.46 & -0.15 & $\alpha_{\rm{s}3}$ & -46.88 & 20.04 & $^{*}$-2.34 \\
$\alpha_{\rm{c}4}$ & -12.54 & 1.53 & $^{*}$-8.19 & $\alpha_{\rm{s}4}$ & -4.75 & 1.61 & $^{*}$-2.95 \\
$\alpha_{\rm{c}5}$ & 95.96 & 41.68 & $^{*}$2.30 & $\alpha_{\rm{s}5}$ & 35.37 & 41.88 & 0.84 \\
$\alpha_{\rm{c}6}$ & -9.17 & 5.14 & -1.78 & $\alpha_{\rm{s}6}$ & -9.70 & 4.93 & $^{*}$-1.97 \\
$\alpha_{\rm{c}7}$ & 71.91 & 35.59 & $^{*}$2.02 & $\alpha_{\rm{s}7}$ & 30.41 & 34.64 & 0.88 \\
$\alpha_{\rm{c}8}$ & -30.56 & 6.18 & $^{*}$-4.95 & $\alpha_{\rm{s}8}$ & 3.95 & 6.10 & 0.65 \\
$\beta_{\rm{c}2}$ & -16.49 & 1.42 & $^{*}$-11.61 & $\beta_{\rm{s}2}$ & -0.36 & 1.67 & -0.21 \\
$\beta_{\rm{c}4}$ & -14.96 & 2.03 & $^{*}$-7.38 & $\beta_{\rm{s}4}$ & -10.02 & 2.19 & $^{*}$-4.58 \\
$\beta_{\rm{c}6}$ & -11.45 & 5.64 & $^{*}$-2.03 & $\beta_{\rm{s}6}$ & 10.08 & 5.88 & 1.71 \\
$\beta_{\rm{c}8}$ & -20.13 & 6.34 & $^{*}$-3.18 & $\beta_{\rm{s}8}$ & -25.09 & 6.48 & $^{*}$-3.87 \\
		\hline
	\end{tabular}
\end{table}

\end{document}